\documentclass[aps,twocolumn,pre,showpacs,eqsecnum]{revtex4-1}
\usepackage{graphpap}
\usepackage[dvips]{graphicx}
\usepackage[dvips]{graphics}
\usepackage{color}

\begin{document}

\title{Charge detection in a bilayer graphene quantum dot}

\author{%
  S. Fringes$^1$,
  C. Volk$^{1,2}$,
  C. Norda$^1$,
  B. Terr\'{e}s$^{1,2}$,
  J. Dauber$^{1,2}$,
  S. Engels$^{1,2}$,
  S. Trellenkamp$^{1,2}$, and
  C. Stampfer$^{1,2}$}

 \affiliation{
$^1$JARA-FIT and II. Institute of Physics B, RWTH Aachen University, 52074 Aachen, Germany, EU\\
$^2$Peter Gr\"unberg Institut (PGI-9), Forschungszentrum J\"ulich, 52425 J\"ulich, Germany, EU
}

\date{ \today}

 \begin{abstract}
We show measurements on a bilayer graphene quantum dot with an integrated charge detector. The focus lies on enabling charge detection with a 30\,nm wide bilayer graphene nanoribbon located approximately 35 nm next to a bilayer graphene quantum dot with an island diameter of about 100\,nm. Local resonances in the nanoribbon can be successfully used to detect individual charging events in the dot even in regimes where the quantum dot Coulomb peaks cannot be measured by conventional techniques. \end{abstract}

 \maketitle

\section{Introduction}
Graphene exhibits interesting electronic properties such as high carrier mobilities and potentially long spin lifetimes which make this material a promising candidate for future quantum devices and quantum information technology in general~\cite{trau07}. In particular single- and few-layer graphene quantum dot devices are expected to feature weak spin orbit interaction and weak hyperfine coupling which is an advantage compared to state-of-the-art GaAs-devices~\cite{min06,her06}. However, the gapless band structure makes it hard to electrostatically confine charge carriers. This can be overcome by nanostructuring graphene, where it has been shown that a disorder-induced energy gap makes carrier confinement possible. Consequently, graphene nanoribbons~\cite{chen07,han07,mol09,stam09,tod09,liu09,gal10,han10,ter11}, single-electron-transistors (SETs)~\cite{sta08,stam08nano,ihn10}, graphene quantum dots (QD)~\cite{gue08,pon08,sch09} even with charge detectors~\cite{gue08,gue11,wang10} and graphene double quantum dots~\cite{mol09a,liu10} have successfully been demonstrated. For gaining better control over the confinement bilayer graphene nanostructures are becoming of increasing interest since they may additionally allow to open a band gap by applying an out-of-plane electric field~\cite{oost08,zha09}. This may reduce the influence of disorder and localized edge states, which are found to limit single-layer graphene nanodevices.\\
Here we present a bilayer graphene QD with an integrated nearby nanoribbon acting as a SET-like charge detector (CD). Charge detection techniques are in general useful to measure ultra-small currents, to detect individual charging events and to read out individual spin-qubit states~\cite{pet05}.
From a fundamental point of view charge detectors have also successfully been used for investigation of detailed electron tunneling statistics and for probing electron-electron correlations~\cite{gus06}.

\section{Device Fabrication}
\begin{figure*}[ht]%
\includegraphics*[keepaspectratio=true,width=.8\textwidth]{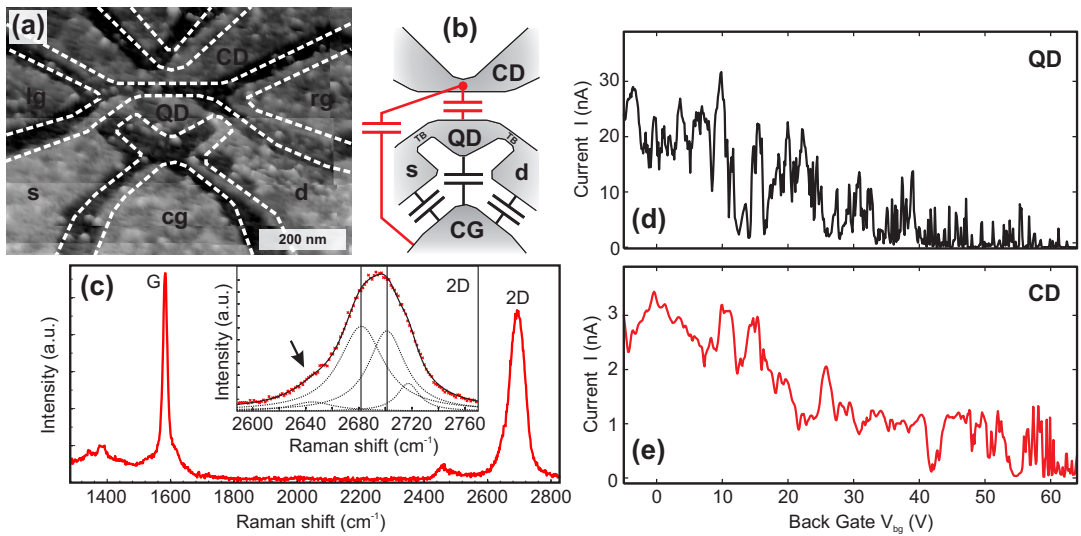}
\caption{%
	(a) Atomic force microscope (AFM) image of the measured device. The dashed lines highlight the contours of the graphene (light gray). 
	(b) Capacitive model for
the analysis of the quantum dot charge detector system.
	(c) Raman spectrum of the measured device. The 2D-peak is highlighted in the inset.
	(d,e) Back gate characteristics of the quantum dot (d) and the charge detector (e) device
	($V_{b,QD}$ = 2\,mV, $V_{b,CD}$ = 0.5\,mV and T $\approx$ 4\,K).
    }
\label{fig1}
\end{figure*}
The devices are fabricated by mechanical exfoliation of natural graphite by adhesive tapes~\cite{nov04,nov05}. The substrate material consists of highly doped silicon (Si$^{++}$) covered with 300\,nm of silicon oxide (SiO$_2$) in order to identify few-layer graphene flakes due to absorption and interference. Before deposition, reference alignment markers are patterned on the substrate to relocate the flakes for further processing and characterization. In particular Raman spectroscopy measurements are used to identify bilayer graphene flakes~\cite{fer06,graf07}.
The graphene flakes have to be structured to submicron
dimensions in order to fulfill the device design requirements.
We use a technique based
on resist spin coating, electron beam lithography (EBL),
development and subsequent etching of the unprotected
graphene. We use an EBL resist
[polymethylmethacrylate (PMMA)] with a thickness of
80\,nm and a short etching time to define the nanoscale structures.
We follow earlier work~\cite{sta08}, where it has been shown that
short (5 to 15\,s) mainly physical reactive ion etching (RIE)
based on argon and oxygen (80:20) provides good results
without influencing the overall quality of the graphene flake~\cite{mol07}. After etching and removing the residual EBL resist, the
graphene nanostructures are contacted by an additional
EBL step, followed by metalization and lift-off. Here we
evaporated 5\,nm chromium (Cr) and 50\,nm gold (Au) to
contact the graphene quantum dot and charge detector devices.
\\
An atomic force microscope image of one of the measured devices is shown in Figure~\ref{fig1}(a), where the graphene structures (bright areas) resting on SiO$_2$ (dark areas) are highlighted by white dashed lines. The smaller bumps increasing the surface roughness are  most likely due to PMMA residues. The bilayer graphene QD device consists of two 25-30\,nm wide and 75\,nm long graphene constrictions separating source (s) and drain (d) contacts from the central quantum dot. The bilayer graphene quantum dot has a diameter of approximately 100\,nm.
The two constrictions and the island are electrostatically tuned by three gates, i.e. the left and right gate (lg and rg) and a central gate (cg) respectively.
The highly doped silicon back gate (bg) additionally allows to adjust the overall Fermi level.
Moreover a 30\,nm wide bilayer graphene nanoribbon is placed approximately 35\,nm next to the island, which acts as a charge detector, as shown below.In Figure~1(b) we illustrate the capacitive model of the device allowing (i) to describe the operation of the coupled graphene device and (ii) to discuss the measurements. 
\\
The Raman spectrum shown in Figure~1(c) proves that the investigated device is indeed made of a bilayer graphene flake. In contrast to single-layer graphene four Lorentzians are needed to successfully fit the Raman 2D-peak (see inset in Figure~1(c)). The FWHM of the 2D-peak is 58.2\,cm$^{-1}$ with a 19.2\,cm$^{-1}$ spacing of the two inner Lorentzian peaks which in combination with the shoulder on the left side (see arrow in the inset of Figure~1(c)) provides a good finger print for the bilayer nature~\cite{fer06,graf07}.

\section{Measurements}
\begin{figure*}[t]
\includegraphics*[keepaspectratio=true,clip,width=.8\textwidth]{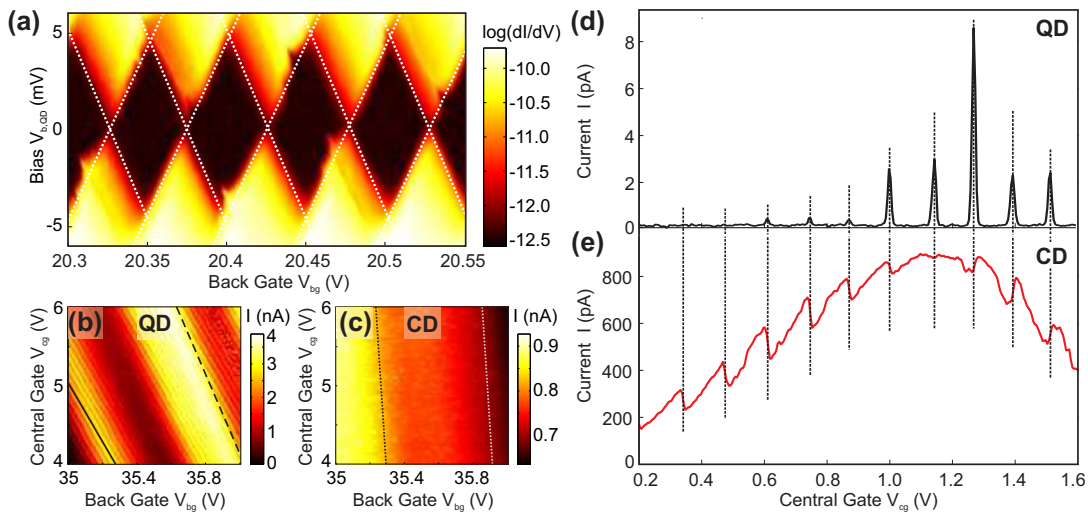}
\caption{%
  (a) Differential conductance $dI_{QD}/dV_{b,QD}$ plotted as a function of the back gate $V_{bg}$ and bias $V_{b,QD}$.
  (b)/(c) Source-drain current through the QD/CD as a function of the back gate $V_{bg}$ and central gate $V_{cg}$ for constant bias $V_{b,QD/CD}$ = 0.5\,mV.
   Current through the (d) quantum dot and through the (e) charge detector $I_{CD}$ as a function of the central gate $V_{cg}$. The back gate is fixed at $V_{bg}$~=~60\,V at a temperature of T~$\approx$~1\,K.   }

\label{fig2}
\end{figure*}
The measurements have been performed in a pumped
$^4$He system with a base temperature of 1.4 K and in a dilution
refrigerator. 
 We have measured the two-terminal conductance
through the bilayer graphene quantum dot and charge detector by applying a symmetric DC bias voltages while measuring the current through the devices with a resolution below 50 fA.

\subsection{Device Characterization}
To characterize the quantum dot and charge detector device on a large energy scale, back gate characteristics are recorded. Figures 1(d) and 1(e) show the source-drain currents as a function of the back gate voltage (at a temperature of approximately 4\,K) for the QD (Fig. 1(d)) and the CD (Fig. 1(e)) devices respectively.
In both cases we observe a so-called transport gap~\cite{stam09} for large back gate values. For the QD we observe the transport gap for $V_{bg} > 40$\,V while for the CD the gap appears at $V_{bg} > 60$\,V (at $V_{b,QD} = V_{b,CD} = 0.5$\,mV). For lower back gate voltages transport is hole-dominated and the different p-doping of the QD and CD is most likely due to atmospheric O$_2$ binding~\cite{ryu10}.
The large-scale current fluctuations in the transport gap region of the QD device are due to local resonances in the graphene constrictions acting as tunnel barriers (TB).
In Figure 2(a) we show Coulomb diamonds, i.e. differential conductance measurements of the dot ($G_{QD}=dI_{QD}/dV_{b,QD}$) as function of bias voltage $V_{b,QD}$ and back gate voltage $V_{bg}$ from a different cool down, where the transport gap was shifted to lower back gate voltages. From the diamond extent in bias direction and the slope of the Coulomb diamond edges a charging energy of $E_{C} = 4.7 \pm 0.1$\,meV and a back gate lever arm of $\alpha_{bg} \approx 0.093$ 
are extracted. \\ 
By focusing on a larger back gate voltage range of 1\,V  and stepping the central gate voltage (shown in Fig. 2(b)) large-scale current fluctuations caused by transmission modulations in the TB and Coulomb blockade resonances in the QD are resolved.
The TB induced resonance can be distinguished from the dot resonances by its
larger width and its different slope given by $\alpha_{cg,TB}$~/~$\alpha_{bg,TB}~= 0.197$ (dashed line)
compared to $\alpha_{cg,QD}$~/~$\alpha_{bg,QD}$~= 0.266 (solid line).
This increased slope is due to
the larger distance of the graphene constrictions (i.e. TB) to the central gate
as compared to the dot-cg distance.
By also measuring such a map for the current through the CD (Fig. 2(c))
we observe that the CD resonances used for detection exhibit an even
steeper slope $\alpha_{cg,CD} / \alpha_{bg,CD} = 0.030$, which is due to the even larger detector-central gate distance.

\subsection{Charge Detection}
For studying the electrostatic coupling of the CD and the QD device the back gate voltage is set to $V_{bg}$ = 60\,V in order to operate both devices in the transport gap regime (see Figs.~1(d,e)). The current through the QD shows clear Coulomb blockade resonances as a function of the central gate voltage $V_{cg}$ ($V_{b,QD}$~=~0.5\,mV; see Figure~2(d)). The almost equal Coulomb peak spacing of $\Delta V_{cg}$ = 129.5\,$\pm$\,7.3\,mV is extracted from 24 charging events. Compared to Figure~2(a) this data are from a different cool down, leading to a different relative lever arm. The different heights of the Coulomb peaks are most likely due to transmission modulations originated by the graphene constrictions acting as TBs~\cite{sta08}.\\
In Figure~2(e) we show the current through the CD which has been measured at the very same time ($V_{b,CD}$ = 0.5\,mV). The resonance of the CD appears much more broadened as a consequence of the smaller lever arm compared to the QD. Furthermore, we observe on top of the peak-shaped CD resonance conductance steps that are well aligned (see dotted lines) with single charging events on the nearby QD. This is a clear signature of the capacitive coupling between the QD and the CD. Whenever an extra electron is added to the dot it changes the electrostatic surrounding of the CD which leads to a sharp step in the CD current. From an analysis of more than 30 charging events, an average shift of the CD resonance in central gate voltage of $\Delta V_{cg,cd}$ = 66.7 $\pm$ 7.9\,mV is obtained (see label in Fig.~2(e)).
These shifts (i.e. current steps) are also visible in regimes where the Coulomb peaks are strongly suppressed and cannot be measured directly by the QD current (see most left dashed lines in Figs.~2(d) and 2(e)).\\
The charge sensitivity of the bilayer graphene charge detector is approx. $3 \cdot 10^{-3}\,\rm{e}/\sqrt{\rm{Hz}}$, where the band with is limited by our low noise amplifiers ($\Delta f = 650$\,Hz).

\subsection{Temperature Dependence}
\begin{figure*}[htb]%
\includegraphics*[keepaspectratio=true,clip,width=.8\textwidth]{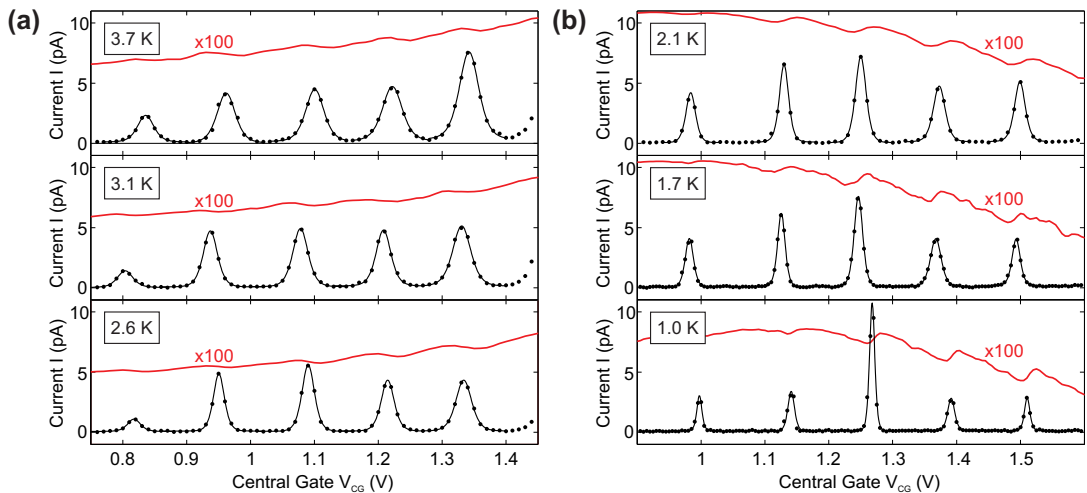}
\caption{ (a),(b) The current through the quantum dot (black) and charge detector (red) is plotted as function of central gate voltage for different temperatures measured at the mixing chamber. For clarity reasons the current of the CD is divided by 100. Additionally an offset of 4\,pA is added in panel (a) for clarity.
}
\label{Fig3}
\end{figure*}
\begin{table*}[ht]
  \caption{Electron temperature extracted from fitting the data shown in Figure 3.}
{\small
\hfill{}
\begin{tabular}{|l||c|c|c|c|c|c|}
\cline{2-7}
\hline
Base Temperature $T_{MC}$ (K)&3.7&3.1&2.6&2.1&1.7&1.0\\
\hline
Electron Temperature $T_{E}$ (K)&(3.76 $\pm$ 0.28)&(3.18 $\pm$ 0.31)&(2.66 $\pm$ 0.36)&(2.13 $\pm$ 0.21)&(1.78 $\pm$ 0.25)&(1.20 $\pm$ 0.18)\\
\hline
\end{tabular}}
\hfill{}
\label{tb:tablename}
\end{table*}
To investigate the influence of the temperature on the charge detector operation we measured the current through the QD and CD as function of the central gate and different mixing chamber temperatures, $T_{MC}$ (see Figure~3). With increasing temperature (see labels in Fig.~3) the broadening of the Coulomb peaks is increasing leading to Coulomb oscillations at elevated temperatures. The charge detection signal, in particular the current steps exhibit also a significant temperature broadening but they are still visible for temperatures up to 4\,K. Moreover it should be noted that the TB transmission is also affected by the increasing temperature. The Coulomb peaks have successfully been fitted by the following relation~\cite{ihn},
\begin{equation}
I(V_{cg},T_E) = \frac{I_0}{ cosh^2 \left[ \alpha_{cg} (V_{cg} - V^{peak}_{cg} ) / 2 k_B T_E \right]},
\label{eq:cosh2}
\end{equation}
where $V^{peak}_{cg}$ is the position of the Coulomb peak and $T_E$ the electron temperature of the system. The average values of the electron temperature are shown in Table~1 for a central gate lever arm of $\alpha_{cg}$ = 0.034, which has been extracted independently.\\

\section{Conclusion}

In summary, we demonstrated the functionality of a bilayer graphene
quantum dot with an integrated charge detector based on a nearby nanoribbon.
We confirm the detection of charging events in
regimes where Coulomb blockade resonances can hardly be
measured or resolved because the current levels are below the current resolution. In contrast to state-of-the-art
quantum-point contact charge detectors, we do not make use of quantized conductance plateaus. We rather use local
resonances in the bilayer graphene nanoribbon to detect individual charging events, very
similar as it has been done in single-layer graphene devices~\cite{gue08}.
This technique is considered to play an important role
for the investigation of future bilayer graphene quantum dots and
in particular double quantum dots, where spin-qubit states may become
accessible.

We thank A. Steffen, R. Lehmann
and U. Wichmann for the help on sample fabrication
and electronics. Discussions with  J. G\"uttinger and S. Dr\"oscher and
support by the JARA Seed Fund and the DFG (SPP-1459 and FOR-912) are gratefully acknowledged.

%
%

\end{document}